\documentclass[showpacs,preprint,aps]{revtex4}
\usepackage{graphicx}
\usepackage{dcolumn}
\usepackage{bm}
\begin{document}
\setcounter{page}{1}
\title
{Quantum reflection and dwell times of metastable states}
\author
{N. G. Kelkar}
\affiliation{ Departamento de Fisica, Universidad de los Andes,
Cra.1E No.18A-10, Santafe de Bogota, Colombia}
\begin{abstract}
The concept of phase and dwell times used 
in tunneling is extended to quantum collisions to derive a relation 
between the phase and dwell time delays in scattering. 
This relation can be used 
to remove the near threshold $s$-wave singularities in the Wigner-Eisenbud
delay times and amounts to an extension of the concept of quantum reflection 
to strong interactions. Dwell time delay emerges as the 
quantity which gives the correct behaviour of the density of states of 
a metastable state at all energies. This fact is demonstrated by investigating 
some recently found metastable states of mesic-nuclei. 
\end{abstract}
\pacs{03.65.-w, 03.65.Nk, 03.65.Xp, 25.70.Ef}
\maketitle

In an attempt to answer the question of how long does a particle take 
to tunnel through a barrier, physicists have given rise to several definitions 
of tunneling times. For example, the dwell time is considered a measure 
of the average time spent by a particle in a given region of space. 
A phase time is defined in terms of the energy derivatives of the phases 
of the reflection and transmission amplitudes.
A traversal time 
was defined by B\"uttiker and Landauer as the interaction time of a 
particle transmitted through a barrier \cite{butikprl}. Though there exist
extensive reviews \cite{hauge,others} on these and some others such
as complex and Larmor times, one does 
find contradictory remarks regarding the physical interpretation of some
of them and the subject in general has remained controversial. It is 
however not the objective of the present work to attempt to resolve any of
the controversies in tunneling and related phenomena like 
the Hartmann effect \cite{harty}. Instead, we start with an 
established relation between the phase time and dwell time which follows
from a derivation in a tunneling problem \cite{hauge,winful} 
and derive a 
similar relation in the context of metastable states in 
quantum collisions.  
The self-interference term appearing in the 
phase time of the 
tunneling problem due to the interference of the reflected and incident 
flux at a barrier re-appears in our derivation 
in terms of a transition matrix in elastic 
scattering. Such a derivation becomes possible only after interpreting 
the reflection appearing 
in tunneling in terms of `quantum reflection' by a strong 
potential in the scattering process.
The dwell time delay expression derived, is useful in extracting 
information on the metastable states 
close to threshold, in contrast to the usual phase time delay (often addressed
simply as `time delay' of Wigner \cite{wigner}) 
which as we shall see becomes singular
in the $s$-wave scattering near threshold.

We illustrate the above relation by applying it 
to the realistic case of the quantum mechanical scattering of 
$\eta$ mesons on light nuclei for the location of unstable eta-mesic nuclear 
states. The attractive nature of the $\eta$-nucleon 
interaction arising due to the proximity of the $\eta$- nucleon 
threshold to the $s$-wave nucleon 
resonance N$^*$(1535), motivated the searches of $\eta$-mesic nuclei. 
The results on the eta-mesic states obtained within the various 
theoretical models of the $\eta$-nucleon interaction are relevant in the 
light of recent experimental claims \cite{pfieff} of such states and 
speculations based on $\eta$-production data \cite{data}. 

We start by considering a relation derived in \cite{winful} in context with 
tunneling through a potential barrier in one dimension (1D) (though the
final results would have relevance for the case of three dimensions (3D), 
as also discussed in \cite{winful}).
For a particle of energy $E = \hbar^2 k^2 /2 \mu$ ($\hbar k$ is the 
momentum), incident on the barrier,  
it was shown in \cite{winful} that, 
\begin{equation}\label{win9new}
\tau_{\phi}(E) \,=\, \tau_D (E) \,-\, \hbar \,\, 
{Im(R) \over k} \,\,{dk \over dE}\,,  
\end{equation}
where $\tau_{\phi}(E)$ which is given in terms of 
a weighted sum of the energy derivative of 
the reflection and transmission phases follows from the  
standard definition \cite{hauge,winful} of phase time. 
The first term on the right hand side is the  
`dwell time' \cite{winful} which is a measure of the time spent by the particle 
in the barrier region regardless
of whether it is ultimately transmitted or reflected.   
The second term on the right is the self-interference term which 
arises due to the overlap of the incident and reflected waves in front of
the barrier \cite{winful}. 
As we shall see later, it is this term which relates 
to threshold singularities in the phase time delay for 3D $s$-wave collisions. 
Eq. (\ref{win9new}) and its 
physical interpretation is discussed extensively in 
\cite{leavens} and 
the case of scattering in 3D in \cite{martin}. 

We note that $\tau_{\phi}$ (in 1D or 3D), is not the 
same as the `time delay' in \cite{wigner,smith}. 
The scattering 
phase shift derivative in Wigner's time delay, 
namely, $\tilde{\tau}_{\phi}(E) = 2 \,\hbar\, d\delta/dE$, 
is related through the Beth-Uhlenbeck 
formula to the density of states \cite{beth}, 
\begin{equation}\label{bethuhl}
\sum_l \,g_l(E)\,-\,g_l^0(E)\,=\, \sum_l \,{2l+1 \over \pi}\, 
{d\delta_l(E) \over dE}\, , 
\end{equation}
where $g_l(E)$ and $g_l^0(E)$ are the densities of states with and without 
interaction respectively. 
In this sense, $\tau_{\phi}(E)$ and $\tau_D(E)$ are the times
spent `with interaction' and not a difference which represents time delay. 
There also exists a general connection between the dwell 
times and the density of states in the barrier region 
for a 3D \cite{iacon} and a 1D system \cite{gaspa}.
The $\tau_D(E)$ is also a measure of the time 
spent in the barrier region and hence is `the density of states with 
interaction'. 
Though both the dwell and phase times are densities of states with interaction,
the interaction in the two cases differs. In the case of dwell time it is
related to the time spent inside the barrier or the time spent by colliding
particles within a certain distance say `r' of each other. In the case of the
phase time, it is the difference between the time of arrival of the wave 
packet at the barrier and the time when it leaves. In this case, the
incident wave has mixed with the reflected one and this affects the time
spent in the barrier. In fact, numerical simulation shows that the traversal
time of a wave packet through a barrier agrees with the phase time rather well 
\cite{collins}.

Subtracting the density of states without interaction for a given region  
\cite{smith} 
(or in other words the density of states of free particles, say 
$\tau^0(E)$) from both sides of (\ref{win9new}), we get, 
\begin{equation}\label{delaydwell}
\tilde{\tau}_{\phi}(E) \,=\, 
\tilde{\tau}_D (E) \,-\, \hbar \,\,{Im(R) \over k} \,\, {dk \over dE} 
\end{equation}
where, $\tilde{\tau}_{\phi}(E)\, =\, \tau_{\phi}(E)\,-\,\tau^0(E) $ and
$\tilde{\tau}_D(E) \,=\,\tau_D(E)\,-\,\tau^0(E)$ 
are now the phase and the 
dwell time delay respectively. In 3D, $\tilde{\tau}_{\phi}(E) = \, 
2 \hbar {d \delta \over dE}\,$ is what is generally referred to as Wigner's 
time delay \cite{wigner}. 
It is the `delay times' which are strongly peaked 
at energy values corresponding to a metastable state \cite{smith}, 
and not simply the times (see \cite{we4} and the references therein).
It would now be right to ask if relation (\ref{delaydwell}) 
could act as a general relation 
for any type of interaction, an 
attractive strong one for example without the existence of a Coulomb 
barrier from which the particle would get reflected. 
To understand this, 
we shall briefly dwell upon the concept of `quantum reflection' below.

Recent research involving ultra cold atoms and molecules has drawn 
great attention 
to the phenomenon of quantum reflection \cite{exptqmref}. 
The term `quantum reflection' 
describes the classically forbidden reflection of a particle in a classically 
allowed region (without classical turning points). 
This could happen 
above potential barriers or at the edge of attractive potential tails
and is often relevant at very low energies. 
The phenomenon of reflection in 1D can be viewed as a back scattering in 3D 
$s$-wave collisions. With no angle dependence of the scattering 
amplitude in the case of $s$-waves, the $s$-wave 3D motion can be viewed as a 
1D motion in the radial coordinate.
The importance of the reflection term in (\ref{delaydwell}) at low energies 
becomes obvious as one
examines the relation between the reflection amplitude, $R$, and the 
scattering matrix, $S$, namely, 
$R = |R| e^{i\phi} = -S = - e^{2i\delta}$, where $\delta$ is the scattering
phase shift (and can in general be complex \cite{arneck}). 
In \cite{arneck}, while discussing the effective range theory 
for quantum reflection amplitudes, it was further pointed out that at   
low energies ($k \rightarrow 0$), $|R|$ approaches unity and the phase 
$\phi \sim \pi - 2 a k$, where $a$ is the scattering length and is related to
the transition matrix, $t$, at zero energy. A similar connection between the 
$S$-matrix and the reflection amplitude 
and its phase has been used in \cite{fried1}.
With $R = -S$ 
and $S$ related to the complex 
transition matrix in scattering as, $S = 1 \,-\, i \mu\,k \,(t_R \,+\, it_I)/\pi$, where 
$t_R$ and $t_I$ are the real and imaginary parts of the $t$-matrix 
respectively and $\mu$ is the reduced mass of the system, we obtain the 
`self-interference' term of (\ref{delaydwell}) in terms of $t$ as, 
\begin{equation}\label{sandr}
- \hbar {Im (R) \over k}\, \,{dk \over dE}\,=\, - \hbar \,\mu\, 
{t_R \over \pi} \,\,{dk \over dE}\,. 
\end{equation}
Replacing the low energy behaviour of the reflection amplitude mentioned 
above ($R\sim e^{i(\pi - 2ak)}$), $dk/dE = \mu/\hbar^2 \,k$ 
and the definition of the complex scattering length 
$a = a_R + i a_I$, 
namely, $t(E=0) = - 2 \pi a/\mu$, we see that 
$- \hbar \,{Im (R) \over k}\, {dk \over dE}\,\, 
{\buildrel k \to 0 \over \simeq}\,\, \, 
{2 \,a_R\, \mu \over \hbar \, k}$ as well as $
- \hbar\, \mu\, {t_R \over \pi} \,{dk \over dE} \,\,\, 
{\buildrel k \to 0 \over \simeq} \,\,\, {2 \,a_R \,\mu \over \hbar \, k}$. 
This is indeed also the threshold singularity present in the definition of 
the time delay in $s$-wave collisions near threshold. 
The real scattering phase
shift for $s$-waves, $\delta \rightarrow k \,a_R$ close to threshold and 
Wigner's time delay, $
\tilde{\tau}_{\phi}(E) \,= \,2 \,\hbar\, {d\delta \over dE} 
\,\,\,{\buildrel k \to 0 \over \simeq} \,\,\,{2 \, a_R \, \mu \over \hbar k}$.
Putting together Eqs (\ref{delaydwell}) and (\ref{sandr}), i.e., identifying 
the reflection amplitude in (\ref{delaydwell}) as a quantum reflection 
amplitude, we can evaluate the 
``dwell time delay", $\tilde{\tau}_D(E)$, once the scattering matrix  
in elastic collisions is known. Thus, in scattering processes, 
\begin{equation}\label{fin}
\tilde{\tau}_D(E) \,=\, \tilde{\tau}_{\phi}(E)\, +\, \hbar \,
\mu\, {t_R \over \pi} \,\,{dk \over dE}\,.
\end{equation}

In general, with the phase shift, 
$\delta \propto k^{2l+1}$ (as given in standard
scattering theory books \cite{joachain}), or rather $\delta \propto 
E^{l+1/2}$ (where $E = E_{CM} \,-\,E_{th}$ and $E_{CM}$ and $E_{th}$ are 
the centre of mass and threshold energies respectively),
the energy derivative $d\delta /dE \propto E^{l-1/2}$. Hence this
derivative which is the phase time delay, blows up for $l=0$, when
$E_{CM} \to E_{th}$.
This makes
locating resonances near threshold difficult using
$\tilde{\tau}_{\phi}(E)$.

With the $S$ matrix given by $S = 1 +2 i T$, with $T = - (\mu k/2 \pi) t$ 
and using the delay 
time in elastic collisions, defined as 
$\tilde{\tau}_{\phi}(E) \,=\, 
\Re e [ -i \hbar (\,S^{-1} {dS / dE}\,)\,]$, we get \cite{we4,we31,we32}, 
\begin{equation}\label{tdsmat}
\tilde{\tau}_{\phi}(E) 
\,=\, {2 \hbar \over A}\,\, \biggl[\,{-\mu \over 2 \pi}\,k\,{dt_R \over dE} \,-\, {\mu^2 \,k^2 \over 
2 \pi^2}\, \biggl (\, t_I\,{dt_R \over dE}\, -\, t_R\,{dt_I \over dE}\,\biggr) \, -\, 
{\mu \over 2 \pi}\,t_R \,{dk\over dE}\,\biggr]\,,
\end{equation}
with $A = 1 \,+\, (2\mu k t_I / \pi)\,+\, (\mu^2 \,k^2 (t_R^2\, + \,t_I^2)/\pi^2) $. For elastic scattering in the absence of inelasticities, 
the factor $A = 1$ and the last term can be identified with the 
self-interference delay given in (\ref{sandr}). In the presence of 
inelasticities, $R \,=\,\eta\,e^{2\,i\,\delta_R}$, where $\delta_R$ is  
the real part of the scattering phase shift and $\eta = e^{-2\,\delta_I}$, 
with $\delta_I$ the imaginary part of the phase shift. 
In the context of the discussion in \cite{winful} above their Eq. (9), 
this would be the case 
of a complex potential with losses, 
where $|R|^2 \,+\,|T|^2$ is not unitary.
One could write $|R|^2\,+\,|T|^2\,=\,1\,-\,|A|^2\, =\,|\xi|^2$, 
where $|A|^2$ is an absorption factor and for a lossless barrier,
$|\xi|^2$ is unity.  
However, due to the fact that time is real, the additional term which arises, 
does not contribute to the delay times or the density of states 
and the general relation between the phase and dwell 
times in (\ref{win9new}) and hence the delay times in (\ref{delaydwell}) 
remains the same. 

We shall now use Eq. (\ref{fin}) to evaluate the dwell time delay distribution 
for the $\eta$-$^3$He and $\eta$-$^4$He systems. 
The model of the t-matrix for $\eta$-$^3$He and $\eta$-$^4$He 
elastic scattering has already been developed in \cite{we31,we32} using few body
equations within the finite rank approximation (FRA) and hence we shall not 
repeat it here. Considering the uncertainty in the knowledge of the 
eta-nucleon ($\eta$ N) 
interaction, in \cite{we31}, these distributions were evaluated
for different models of the elementary $\eta$ N interaction. 
Since the nucleus in $\eta$-nucleus scattering remains in 
its ground state in the intermediate state in the FRA of few body equations,
we shall focus on the near threshold region. Besides, the 
$\eta$-mesic states of interest are also expected to be close to threshold,  
according to some recent experimental claims \cite{pfieff}.  
One of the objectives 
of these calculations is also to demonstrate the extension of the concept
of ``quantum reflection in nuclear physics". To the best of our knowledge, 
quantum reflection phenomena have been observed \cite{exptqmref} in the case 
of long ranged attractive potentials. With the absence of a Coulomb `barrier' 
for the neutral $\eta$ mesons, one can understand the singular term related
to the reflection amplitude only in terms of a `quantum reflection'.

In Fig. 1, the delay times in the elastic scattering reaction, $\eta\,^3$He 
$\rightarrow \eta\,^3$He are shown as a function of the energy, 
$E\, =\, E_{\eta^3He}\,-\,m_{\eta}\,-\, m_{^3He}$, where $E_{\eta^3He}$ is 
the total energy available in the centre of mass of the $\eta\,^3$He system. 
The solid lines correspond to the Wigner-Eisenbud time delay evaluated 
using (\ref{tdsmat}). The time delay at negative energies 
(see \cite{we31,we32} for the physical interpretation) corresponds to 
$k \to ik$ (hence $E = -k^2/2\mu$) and is useful in locating the bound 
and quasibound states with negative binding energy. In \cite{we31} the
correctness of the method for bound states has been demonstrated 
with the example of the lowest stable nucleus, namely, the deuteron. 
The one-to-one correspondence between the poles of the $S$-matrix and the 
peaks in the delay times has been shown for the case of quasi-virtual and 
quasi-bound states of the $\eta$-meson and the deuteron in \cite{we32}. 
This correspondence holds at all energies for partial waves with $l\ne 0$ and 
for $l=0$ only at energies quite away from threshold. 
At negative energies, in the interference term in (\ref{fin}), $t_R$ 
gets replaced by $t_I$ and $a_R$ by $a_I$ in the $k\to 0$ limit. 
The dashed curves in Fig. 1 are the self-interference 
delays due to quantum reflection as given 
in (\ref{sandr}). The shaded regions correspond to the dwell time 
delay which is obtained after subtracting the self-interference 
term from the delay time $\tilde{\tau}_{\phi} (E)$ as in (\ref{delaydwell}). 
The three different plots correspond to three distinct models of 
the $\eta$ N interaction which lead to the three different 
$\eta$ N scattering lengths, namely, $a_{\eta N} = (0.88,0.41)$ fm \cite{fix}, 
$a_{\eta N} = (0.28,0.19)$ fm \cite{bhalerao} and 
$a_{\eta N} = (0.51,0.26)$ fm \cite{green}. A peak just
above threshold (a resonance of $\eta\,^3$He around $0.5$ MeV excitation 
energy) appears for the largest of the scattering lengths. 
The dwell time delay in this case shows a clear Lorentzian distribution. 
For the smaller scattering lengths, the peaks become broader and spread over 
to negative energies. For $a_{\eta N} = (0.28,0.19)$ fm, 
a broad peak appears around $-5$ MeV, in 
agreement with the experimental findings \cite{pfieff} of $\eta$-mesic
helium.  
\begin{figure}[h]
\includegraphics[width=9cm,height=7cm]{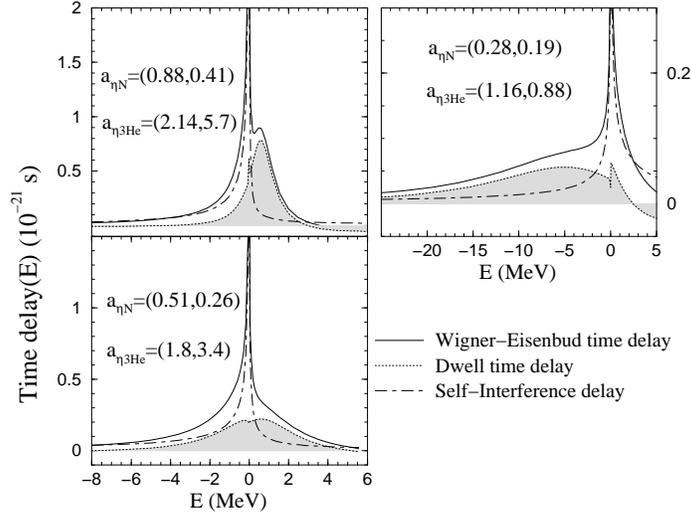}
\caption{\label{fig:eps1} 
Phase time delay (solid lines), 
dwell time delay (shaded region) and the self-interference delay 
(dashed lines) in the reaction,
$\eta \,^3$He $\rightarrow \eta \,^3$He, versus 
$E = E_{\eta \,^3He} - m_{\eta} - m_{^3He}$, for 
models of the $\eta$N interaction with 
three different $\eta$N and $\eta \,^3$He
scattering lengths, $a_{\eta N}$ and $a_{\eta 3He}$, in fm.}
\end{figure}

\begin{figure}[h]
\includegraphics[width=6cm,height=7cm]{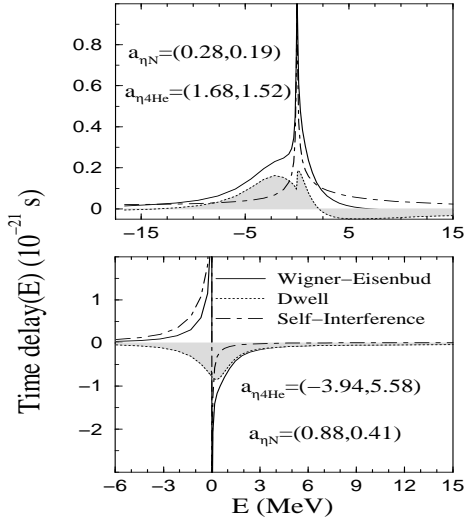}
\caption{\label{fig:eps2} Same as Fig. 1 but for the case of 
$\eta \,^4$He elastic scattering.}
\end{figure}
In Fig. 2 we show the dwell and phase time delay in $\eta \,^4$He elastic
scattering for two $\eta$ N models \cite{fix, bhalerao}. 
Once again, for the smaller scattering length, there is a 
bump at a negative energy of $-2$ MeV. The model with the larger 
scattering length however displays a clear `time advancement' 
in the distribution. Such a negative time delay can be interpreted
in terms of a repulsive $\eta \,^4$He interaction \cite{smith,me1}. 
This means that the dwell time without interaction 
is bigger than the one with interaction, thus also implying that the 
interaction must be repulsive (which is also apparent from the sign of
$\Re e \,[a_{\eta \,4He}]\,=\,-3.94$ fm).

In summary, we began with an expression 
connecting the phase and dwell times obtained in connection with 
quantum tunneling and wrote down the corresponding expression for the
phase and dwell time `delays' (\ref{delaydwell}). 
The relation emerges after noting that the
`phase time delay' of Wigner is simply the difference 
of the density of states with and without interaction as given by the
Beth-Uhlenbeck formula. The `dwell time delay' is also a difference
of the density of states, however, it is free of the threshold singularity 
present in the definition of the $s$-wave phase time delay.
The self-interference term 
with a reflection amplitude in tunneling can be identified as a term
given in terms of the $t$-matrix in scattering only 
by introducing the concept of `quantum reflection'. Since such a 
self-interference delay term in scattering depends on the $t$-matrix, 
it is not necessary to have a barrier, like a Coulomb barrier for example
through which the $\alpha$ particle tunnels in the case of $\alpha$ decay. 
That the 
dwell time delay gives the correct energy behaviour of metastable states is 
demonstrated by applying it to the $s$-wave scattering of the neutral 
$\eta$ mesons and helium nuclei. 
The practical importance of the relation lies in the separation of the 
$s$-wave threshold singularity which can be confused with a zero energy
resonance.   
Lorentzian peaks in the
dwell time delay distributions, which have some support from experimental
data on $\eta$-mesic nuclear states emerge after the subtraction of the 
self-interference delay from the phase time delay. This method can in
general be applied to locate $s$-wave unstable states near threshold.


\begin{thebibliography}{99}
\bibitem{butikprl}
M. B\"uttiker and R. Landauer, Phys. Rev. Lett. {\bf 49}, 1739 (1982);
{\it ibid}, Phys. Scr. {\bf 32}, 429 (1985).
\bibitem{hauge}
E. H. Hauge and J. A. St\o vneng, Rev. Mod. Phys. {\bf 61}, 917 (1989); 
E. H. Hauge, J. P. Falck and T. A. Fjeldly, Phys. Rev. {\bf B 36}, 
4203 (1987).
\bibitem{others}
J. G. Muga, R. Sala-Mayato and I. L. Egusquiza, {\it Time in 
Quantum Mechanics}, New York: Springer (2002); R. Landauer and Th. Martin, 
Rev. Mod. Phys. {\bf 66}, 217 (1994);  
V. S. Olkhovsky and E. Recami, Phys. Rep. {\bf 214}, 339 (1992); 
H. Winful, Phys. Rep. {\bf 436}, 1 (2006).
\bibitem{harty}
J. G. Muga, I. L. Egusquiza, J. A. Damborenea and F. Delgado, Phys. Rev. 
{\bf A 66}, 042115 (2002); T. E. Hartman, J. Appl. Phys. {\bf 33}, 
3427 (1962).
\bibitem{winful}
H. G. Winful, Phys. Rev. Lett. {\bf 91}, 260401 (2003).
\bibitem{wigner}
E. P Wigner, Phys. Rev. {\bf 98}, 145 (1955); E. P. Wigner and L. Eisenbud
Phys. Rev. {\bf 72}, 29 (1947).
\bibitem{pfieff}
M. Pfeiffer {\it et al}., Phys. Rev. Lett. {\bf 92}, 252001 (2004).
\bibitem{data}
B. Mayer {\it et al}., Phys. Rev. {\bf C 53}, 2068 (1996).
\bibitem{leavens}
C. R. Leavens and G. C. Aers, Phys. Rev. {\bf B 39}, 1202 (1989); 
{\it ibid}, Solid State Commun. {\bf 63}, 1107 (1987); 
M. B\"uttiker, Phys. Rev. {B 27}, 6178 (1983);  
\bibitem{martin}
Ph. A. Martin, Acta Phys. Austrica Suppl. {\bf 23}, 157 (1981).
\bibitem{smith}
F. T. Smith, Phys. Rev. {\bf 118}, 349 (1960).
\bibitem{beth}
E. Beth and G. E. Uhlenbeck, Physics {\bf 4}, 915 (1937); 
K. Huang, {\it Statistical Mechanics} (Wiley, New York, 1963).
\bibitem{iacon}
G. Iannaccone, Phys. Rev. {\bf B 51}, R4727 (1995).
\bibitem{gaspa}
V. Gasparian and M. Pollak, Phys. Rev. {\bf B 47}, 2038 (1993). 
\bibitem{collins}
S. Collins, D. Lowe and J. R. Barker, J. Phys. C: Solid State
Phys. {\bf 20}, 6213 (1987).
\bibitem{we4}
N. G. Kelkar, M. Nowakowski, K. P. Khemchandani
and S. R. Jain, Nucl. Phys. {\bf A730}, 121 (2004); 
N. G. Kelkar, M. Nowakowski and K. P. Khemchandani, 
Mod. Phys. Lett. {\bf A 19}, 2001 (2004); 
{\it ibid}, Nucl. Phys. {\bf A 724}, 357 (2003); {\it ibid}, 
J. Phys. {\bf G 29}, 1001 (2003). 
\bibitem{exptqmref}
F. Shimizu, Phys. Rev. Lett. {\bf 86}, 987 (2001); 
T. Pasquini {\it et al}., Phys. Rev. Lett. {\bf 93}, 223201 (2004); 
{\it ibid} {\bf 97}, 093201 (2006); H. Oberst, Phys. Rev. {\bf A 71}, 
052901 (2005).
\bibitem{arneck}
F. Arnecke, H. Friedrich and J. Madro\~nero, Phys. Rev. {\bf A 74}, 
062702 (2006).
\bibitem{fried1}
H. Friedrich and A. Jurisch, Phys. Rev. Lett. {\bf 92}, 103202 (2004); 
E. Kogan, P. A. Mello and He Liqun, Phys. Rev. {\bf E 61}, R17 (2000); 
U. Kuhl, M. Mart\'inez-Mares, R. A. M\'endez-S\'anchez and H. -J. St\"ockmann, 
Phys. Rev. Lett. {\bf 94}, 144101 (2005); 
H. Friedrich and J. Trost, Phys. Rep. {\bf 397}, 359 (2004).
\bibitem{joachain} 
C. J. Joachain, {\it Quantum Collision Theory} (North-Holland, 1975). 
\bibitem{we31}
N. G. Kelkar, K. P. Khemchandani and B. K. Jain,
J. Phys. {\bf G 32}, L19 (2006).
\bibitem{we32}
N. G. Kelkar, K. P. Khemchandani and B. K. Jain, J. Phys. {\bf G 32}, 1157 
(2006). 
\bibitem{fix}
A. Fix and H. Arenh\"ovel, Eur. Phys. J A {\bf 9}, 119 (2000);
{\it ibid}, Nucl. Phys. A {\bf 697}, 277 (2002).
\bibitem{bhalerao}
R. S. Bhalerao and L. C. Liu, Phys. Rev. Lett. {\bf 54} (1985) 865.
\bibitem{green}
A. M. Green and S. Wycech, Phys. Rev. C {\bf 71}, 014001 (2005).
\bibitem{me1}
N. G. Kelkar, J. Phys. G: Nucl. Part. Phys. {\bf 29}, L1 (2003).
\end{thebibliography}
\end{document}